\documentclass[aps,prb,reprint,groupedaddress, floatfix]{revtex4-1}
\usepackage{graphicx} 
\usepackage{amssymb} 
\usepackage{amsmath} 
\usepackage{comment}

\let\baraccent=\= 
\renewcommand{\=}[1]{\stackrel{#1}{=}} 

\begin{document}  
  
\title{  
Ultrafast  photocurrent and absorption microscopy of few-layer TMD devices isolate   rate-limiting dynamics driving fast and efficient  photoresponse\\ }
\author{Kyle T. Vogt$^1$,  Su-Fei Shi$^{2}$, Feng Wang$^{3}$ and Matt W. Graham$^{1}$}
\affiliation{$^{1.}$Department of Physics, Oregon State University, Corvallis, OR 97331 USA}
\affiliation{$^{2.}$Department of Chemical Engineering, Rensselaer Polytechnic Institute, Troy, NY 12180 USA}
\affiliation{$^{3.}$Department of Physics, University of California, Berkeley, CA 94720 USA}
\begin{abstract}
Despite inherently poor interlayer conductivity, photodetectors made from few-layer devices of 2D transition metal dichalcogenides (TMDs) such as  WSe$_2$ and MoS$_2$ can still yield a desirably fast ($\leq$90 ps) and efficient ($\epsilon$$>$40\%) photoresponse.  By combining ultrafast photocurrent (U-PC) and transient absorption (TA) microscopy,   the competing electronic escape and recombination rates  are unambiguously identified in otherwise complex  kinetics. Both the  U-PC and TA response of WSe$_2$ yield matching interlayer electronic escape times  that accelerate from 1.6 ns  to 86 ns with applied $E$-field  to predict the maximum device PC-efficiency realized of $\sim$44\%. The slope of  the escape rates versus    $E$-field  suggests out-of-plane electron and hole mobilities of  0.129 and 0.031 cm$^2$/V$s$ respectively. Above $\sim$10$^{11}$ photons/cm$^{2}$ incident flux, defect-assisted Auger scattering greatly decreases efficiency by trapping carriers at vacancy defects. Both TA and PC spectra identify a metal-vacancy sub-gap peak with $\sim$5.6 ns lifetime as a primary trap capturing carriers as they hop between layers. Synchronous TA and U-PC microscopy show  the\   net PC collected is modelled by a kinetic rate-law of  electronic escape competing against  the linear and nonlinear Auger recombination rates. This  simple rate-model  further   predicts the  PC-based   dynamics, nonlinear amplitude and efficiency, $\epsilon$  over a  10$^5$ range of  incident photon flux in few-layer   WSe$_2$ and MoS$_2$  devices.

                                                                           
\end{abstract}
\maketitle 
\begin{figure}[htbp]
        \includegraphics[height=6.4in.]{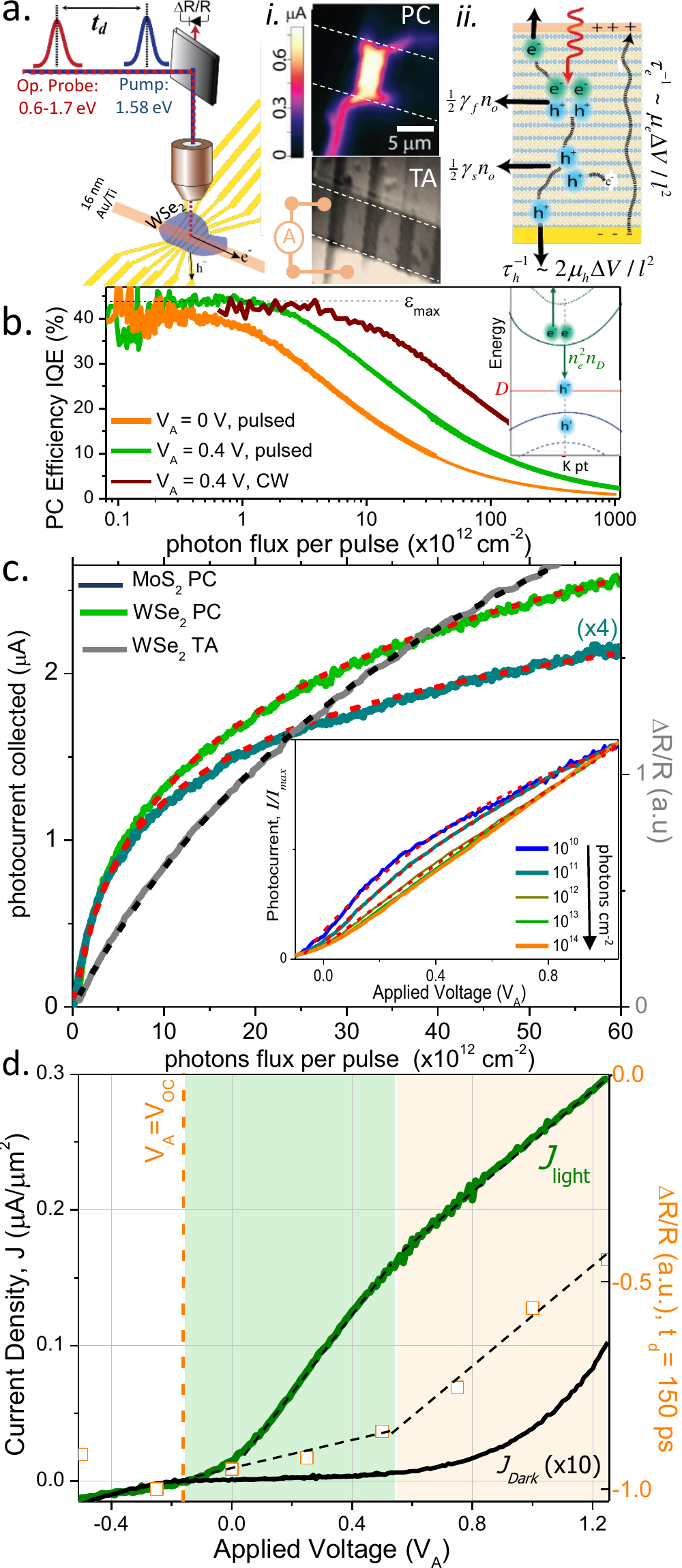}   
        \caption{ \textbf{Nonlinear TMD photocurrents.} \textbf{(a)} Experimental  setup  measuring ultrafast photocurrent (U-PC, $\Delta I(t,V_A)$) and transient absorption ($TA, \Delta R(t,V_A)$) response for  few-layer WSe$_2$ devices shown.   \textbf{i.} U-PC (top) and TA (bottom) spatial map. \textbf{ii.} Device cross-section show reduction of PC and photoconduction (arrows) by fast ($\gamma_f$) and slow ($\gamma$) Auger recombination to vacancy traps.  \textbf{(b) \textbf{\textbf{}}}PC-IQE decreases with photon flux for both pulsed and \textit{CW}-illumination. (\textit{inset}) Auger scattering to defects (\textit{D}) increase with electronic density. {\textbf{(c)}}  PC power dependence for WSe$_2$ (\textit{green}, $0.4$ V) and MoS$_2$ (\textit{blue}, $0.5$ V) fits to Eq. 4 (\textit{red dashed lines}).  TA amplitude (\textit{gray}) at $t_d=0.1$ ps both fit to Eq. 2 (\textit{dashed black line}). (\textit{inset})  Normalized PC $V_A$-sweeps show the nonlinear response is fit \textit{\textit{\textit{red dashed lines} }} by Eqs. 4-5.  \textbf{(d)} Resonant  excited, $J_{light}$ (\textit{green}) and $J_{dark}$ (\textit{black}) current density $V_A$ sweeps   plotted with TA amplitude for $t_d=$150 ps (\textit{orange squares}, \textit{right-axis}). Shading  denotes two quasi-linear regimes associated with PC (\textit{green}) and photoconduction (\textit{orange}). }
\end{figure}

Photodetectors made from few-layer stacked, semiconducting 2D transition metal dichalcogenides (TMDs) like WSe$_2$ and MoS$_2$ can be remarkably fast (sub $\sim$100 ps) and efficient with  internal quantum efficiencies (IQE, $\epsilon$) of photocurrent (PC) collected exceeding  $40$\%.\cite{Yu2013, Massicotte2015,Lee2014,huang_lateral_2014}   In Fig. 1a, our photodetector works by employing a parallel plate sandwich device geometry with a thin top-contact that uses  either  the  built-in or applied voltage ($V_A$)  to collects photoexcited electrons or holes.  Stacked layers of WSe$_2$ or  MoS$_2$  are only weakly coupled by van der Waals forces, resulting in very poor out-of-plane conductivity. To reconcile such weak interlayer coupling with the high PC-IQE and fast photoresponse seen in similar TMD devices, this work isolates the competing kinetics pathways of carrier recombination and interlayer escape in few-layer stacked TMD devices. \cite{Yu2013, Massicotte2015,Vogt:16,zhou_tunneling_2018}

Measuring the electronic escape  rates  driving efficient PC collection remains
 challenging owing to  complex intraband dynamics and  defect-assisted Auger-scattering  in  stacked  semiconductors  like few-layer TMDs.\cite{zipfel_exciton_2019,selig_ultrafast_2019,zhao_evolution_2013} Presently, there exists no reliable  \textit{in-situ} time-resolved method that selectively isolates both the effective linear recombination ($\tau_l^{-1}$) and  electronic escape ($ \tau_{e/h}^{-1}$ ) rates in the ultrafast regime.  Transport based measurements lack the required  time-resolution, while purely optical ultrafast measurements  provide a convoluted weighted-average of all  dynamics, offering no selectivity for  rate-components that are relevant to PC production.\cite{Mueller2018} The high PC-IQE and  nonlinear relaxation kinetics associated with  TMDs give a  nonlinear reduction in PC     that can be time-resolved directly alongside purely optical  measurements like transient absorption (TA) to isolate carrier extraction mechanisms and rates. 

Even under low incident flux and continuous wave (CW) excitation, the interlayer electronic dynamics in TMDs are often dominated by  defect-assisted Auger scattering that localize photocarriers in sub-gap defect sites.\cite{Wang2015a,strait_high_2014,moody_exciton_2016}. For example,   one defect-assisted  Auger recombination   process is depicted in Fig. 1b(\textit{inset}) by showing  electron-electron scattering to  sub-gap defect states with defect density, $n_D$.  Such interactions  have a rate coefficient, $\gamma$  and  a kinetic rate,  $R$$\propto n^2n_D$ that scales quadratically with electron or hole density, $n$.    Prior  optical ultrafast studies show fast and slow Auger rate laws  dominate the kinetic relaxation in few-layer TMDs over the range of photon fluxes employed.\cite{Yuan2015,li_auger_2018,yu_fundamental_2016}  However at the lower photon fluxes practically relevant to  TMD device operation, more complex kinetic rate models are proposed that fully describe the fast electronic scattering and thermalization  within the  indirect-gapped band structures of WSe$_2$ and MoS$_2$.\cite{he_competition_2018,Wang2015,Wang2015b}  While such complex rate laws are now required to rationalize the multi-component relaxations observed in optical TA-based measurements, it remains unclear which kinetic processes are practically relevant to PC-IQE and overall response  times.\cite{merkl_ultrafast_2019, moody_exciton_2016}  As shown in Fig. 1a, we will analytically test  the adequacy of   simpler kinetic rate  models to identify which dynamics processes  sufficiently  predict the net PC-collected and response time in few-layer TMD-photodetectors.

 By combining  ultrafast photocurrent (U-PC) with transient absorption (TA) microscopy,    this work  identifies   bottleneck-kinetic rate mechanisms that intrinsically limit the maximum PC-IQE achievable in few-layer TMD photodetectors. The integrated  U-PC device response is thought to be selective for rate-limiting electronic dynamics  forming the primary bottleneck for PC extraction.\cite{Graham2013, Gabor2012, Massicotte2015,Vogt:16}  However, this assumption is untested and no robustly-used response-function for extracting first-principle kinetics from U-PC kinetics yet exists. Consequently, we employ synchronous \textit{E}-field-dependent TA microscopy, to show how both the \textit{pure-optical} and \textit{purely-electronic} detection regimes independently predict the same high PC-generation efficiency and fast electronic escape times in TMD-based devices.
\\ \\
\textbf{Experimental Analysis I: \textit{Modeling the nonlinear photoresponse of TMDs}.}  
\indent We began by fabricating photoconductive devices illustrated in Fig. 1a of few-layer WSe$_2$ or MoS$_2$ TMDs (\textit{see method details}). The photocurrent  signal was collected from both WSe$_2$ and MoS$_2$ devices by raster scanning a pulsed (160 fs) or continuous wave (CW), diffraction-limited  spot resonant with the lowest, \textit{K}-point transition.  Optically translucent (See supplement S.1) metallic 16/2 nm Au/Ti  layers formed excellent top-contact with a  subset of devices fabricated.  Each device shown in Fig. 1a  is further divided into 6 distinct contact regions.   Only devices generating  a uniform strong PC   signal such as shown in Fig. 1a\textit{i}. were used for this study to help decouple contact and barrier issues.   In Fig. 1a\textit{i.} scanning PC    microscopy spatial maps of a 69 layer WSe$_2$ device shows strong, uniform PC over the sandwiched regions that is 2-3 orders of magnitude greater than any edge-PC signals around the peripheral leads. The corresponding TA scanning microcopy map shown in Fig. 1a\textit{i} is sensitive to the carrier population remaining after at pulse delay time, $t_d =$ 1 ps between pump and pulses resonant with the optical gap.  Using these scanning maps, a 0.9-1.4 $\mu$m laser spot size was localized on the device center for all    PC and TA measurements. As depicted in Fig. 1a\textit{ii} (\textit{dashed arrows}),  TMDs placed between capacitively matched contacts have two modes of operation; i.e. either the photoexcited electron-hole pairs are  extracted (low applied voltage,  $V_A$) as a PC, or a photoconductive current between contacts is optically induced as the diode-like response turns-on at higher at higher $V_A$.\\ 

\indent Fig. 1b shows that as the incident photon flux per pulse increases by  $\sim$10$^4$,  the WSe$_2$  device PC-IQE, $\epsilon \cong n_{PC}/n_{abs}$ drops precipitously (\textit{see supplement \textit{S.1}}  for IQE and EQE calculations). This drop suggests  nonlinear rate kinetics are required to predict the PC amplitude above a $\sim10^{11}$ photons/cm$^2$ threshold.  This  threshold photon flux  in PC-IQE further increases by a factor of $\sim$10  going from the unbiased (\textit{orange line}) to  biased (0.4 V, \textit{green line}), and again to continuous wave (CW) illumination (0.4 V, \textit{red line}).\cite{Massicotte2015, Yu2013}  This shift is understood by noting the   Auger recombination rate, $R\propto n^2n_D$, will increase  both with carrier density (i.e. pulsed vs. CW) and as electron drift velocity slows when $V_A \rightarrow 0$ to enhance carriers present in the device.      While the nonlinear PC-dependence in Fig. 1b-c has been previously  attributed to Auger recombination processes, there exists no predictive analytic model linking the PC and optical responses widely reported in such TMD materials.\\ 
\indent  In agreement with the early few-layer TMD device studies by Yu \textit{et al}.\cite{Yu2013}, Figs. 1b-c show a  highly nonlinear PC  response for resonant  CW and pulsed excitation of both WSe$_2$ and MoS$_2$ devices. This motivates the inclusion of  slow defect-assisted scattering Auger processes in our kinetic model for  PC.  Wang \textit{et al.} show the nonlinear response has contributions from both   fast  Auger recombination with rate, $\gamma_fn$ and slower interlayer defect-assisted Auger recombination processes with rate  $\gamma n$, where $n$ is either the electron or hole density\cite{Wang2015a,vialla_tuning_2019}.
 Fast Auger  processes occurs on a timescale commensurate with the interlayer electron-hole separation time, $\tau_d$ (\textit{see discussion}).\cite{Poellmann2015}  The rate law for slower defect-assisted Auger processes is written as   $R=\gamma n^2$ where $\gamma=n_D\gamma'$, and scales linearly with the constant metal or dichalcogenide vacancy defect concentration.   \\ 
\indent  To model the nonlinear PC response in few-layer  WSe$_2$
 and MoS$_2$ photodetectors, we first consider a simplified kinetic rate law of  the PC-efficiency-limiting rate processes. We approximate the total linear rate, $\tau_{l}^{-1} $ involved in PC generation as the sum of the applied  field-dependent electron/hole net  escape rate, $\tau_{e/h}^{-1}$ and  linear  recombination rate, $   \tau_{r}^{-1}$. This approximation  ignores explicit spatial and valley dependence. Collectively after electron-hole separation,  the rate law for the time-evolution of  electron or  hole photocarriers density $n(t)$ is simply,
\begin{align}
\frac{dn}{dt} & =n_{o} \delta(t) - \frac{n}{\tau_{l}}-\frac{1}{2}\gamma n^{2},\\ n(t,n_o)      & = n_o \left[ {e^{t/\tau_{l}}+ \frac{\tau_l \gamma n_o}{2}   \left(e^{t/\tau_l}-1 \right)} \right] ^{-1}
\end{align}

\noindent The above solution to the rate  law predicts that carrier density grows nonlinearly with initial photocarrier density, $n_o$.\cite{Yuan2017}   In the low carrier density limit ($\gamma n_o \cong 0$),  Eq. 2 gives  the carrier density, $n(t) $    growing linearly with initial carrier density $n_o$  (and incident photon flux). This purely linear PC-response region is clearly seen by the flat-region of Fig. 1b (and later again in Fig. 3b on a log-log scaling).     Fig. 1c plots TA deferential reflectivity   power dependence   for the WSe$_2$(\textit{gray} \textit{line}, right axis) and fits well to Eq. 2 (\textit{dashed black line}). 

Unless otherwise specified, the PC and TA signals are collected at resonant excitation at optical gap $K_A$ of WSe$_2$\ , and both signals evolve from a linear to nonlinear scaling with increasing photon incident flux (see Fig. 1b-c). To test if Eq. 1 predicts the  interlayer PC response in TMDs,  Fig. 1c  (and 3b) we plots the integrated PC response over a 10$^5$ change in photon flux.  The integrated photocurrent density,  $J_{e/h}=-D_{e/h}\frac{dn}{dx}$ of electrons or holes is a spatial transport process traditionally defined through the diffusion coefficient $D_{e/h}$, and any driving applied voltage.\cite{zipfel_exciton_2019} Alternatively, for few-layer devices we instead approximate $J_{e/h}$ in the time-domain by integrating time-dependent carrier density, $n(t)$ given by in Eq. 2 and doing  a time averaged over the mean electronic escape time, $\tau_{e/h}$ to get,     

\begin{align}
J_{e/h}(n_{o})&=\frac{-e}{\tau_{e/h}}  \int_0^{\infty} n(t) dt\\
&=  \frac{-e}{\tau_{e/h}  \tau_{l}\gamma}\ln\left[1+\tau_l\gamma n_o/2\right].
\end{align}
This yields a simple, but highly predictive expression showing  how the PC collected depends nonlinearly on the initial photoexcited carriers created ($n_o$).  In Fig. 1c (and again later in Fig. 3b), the \textit{red-dashed line\textit{}} fits to Eq. 4 show excellent agreement with  PC spanning five orders of magnitude of  photon-flux dependence  for both TMD devices of    56 nm in WSe$_2$ (\textit{green line}) or  43 nm in MoS$_2$ (\textit{blue line}).  The highly predictive nature of Eq. 2 and 4  suggest the simple rate law in Eq. 1 might  sufficiently capture the rate-limiting kinetics germane to PC collection.  This assertion will be further tested  by ultrafast time-resolved PC and TA in section II. 

For photogenerated carriers, the escapes rates have been previously approximated using the drift velocity, $v_D=-\mu_{e/h}E \cong l/\tau_{e/h}$ where $\mu_{e/h}$ is the out-of-plane electron or hole mobility, and $l$ is the mean distance to escape.\cite{Massicotte2015,Gabor2012} In the thin sandwich device capacitive geometry shown in Fig. 1a,  the  perpendicular \textit{E}-field is determined as $E\sim|V_A-V_{OC}|/L=\Delta V/L
$ (\textit{see supplementary section} for justification), where $V_{OC}$ is the device open-circuit voltage. Solving for the electronic escape time, one obtains,

\begin{align}
\tau_{e/h}^{-1} (V_{A}) = \frac{\mu_{e/h}|V_{A}-V_{oc}|}{lL}
\end{align}
  where $l$ is the mean distance of photocarrier transport to the contact Ti/Au contacts.  AFM of the WSe$_2$ device in Fig. 1a\textit{i}. shows the total TMD layer thickness is $ \textit{L}$=56 nm.     In Fig. 1c (\textit{inset}), the combination of  Eqs. 4 and 5, also simulates how the  PC data scales with $V_A$.   Over the 10$^4$ range of photon flux shown, both the linear scaling, $J\propto |V_A-V_{oc}$\textbar, and the nonlinear scaling regions  the $V_A$ dependent PC-response are also modelled well by combining  Eq. 4 and 5 (\textit{red line fits}).

 The green line in Fig. 1d is   the illuminated current density, $J_{light}$ and shows an inflective change at 0.55 V.   The dark current, $J_{dark}$ is $\sim$ 100x smaller than $J_{light}$ and exhibits diode-curve-like behavior for low $V_A$.  As PC-signals shown are detected by using lock-in detection, the weak,  $J_{dark}$ response is removed from all our signals to isolate only the photoresponse. There are two quasi-linear regimes in Fig. 1d \textit{(dashed black line fit})  delineated by the green or orange background shading.  Each region corresponds to PC dominated by photo-extraction or photoconduction, respectively. To compare PC to the purely optical response, on the right-axis of Fig. 1d, the \textit{orange squares} show the amplitude of the TA differential reflectivity signal at a $t_d=$160 ps.   Interestingly, the purely optical TA signal also shows two quasi-linear regions, tracking the PC electronic signal closely.

 When photoexcited carriers are directly extracted as a PC, electric field simulations suggest the mean escape distance is  approximately $l\cong L/2$.\cite{Massicotte2015} However for a  photoconduction  response, the electronic carriers must instead hop the device thickness $L$ between all layers.  In Fig. 1d,  the  PC-response evolves to a photoconductive dominated response as the applied voltage, $V_A$ is swept from the open-circuity  value, $V_{OC}$=-0.089  to 1.30 V. Revisiting   the inset of Fig. 1c, one observes this  evolving inflection-point in photocurrent density  develops for photon flux below $\sim5\times 10^{12}$ photon/cm$^2$ where PC-collection becomes favorable over photoconduction.  In the next section,  the associated dynamic rates driving  the complex PC  response of Fig. 1 are obtained using a novel ultrafast time-domain  analysis of  \textit{on-chip} PC and  TA microscopy.  \\

  \begin{figure}[htbp!]
        \includegraphics[height=7 in.]{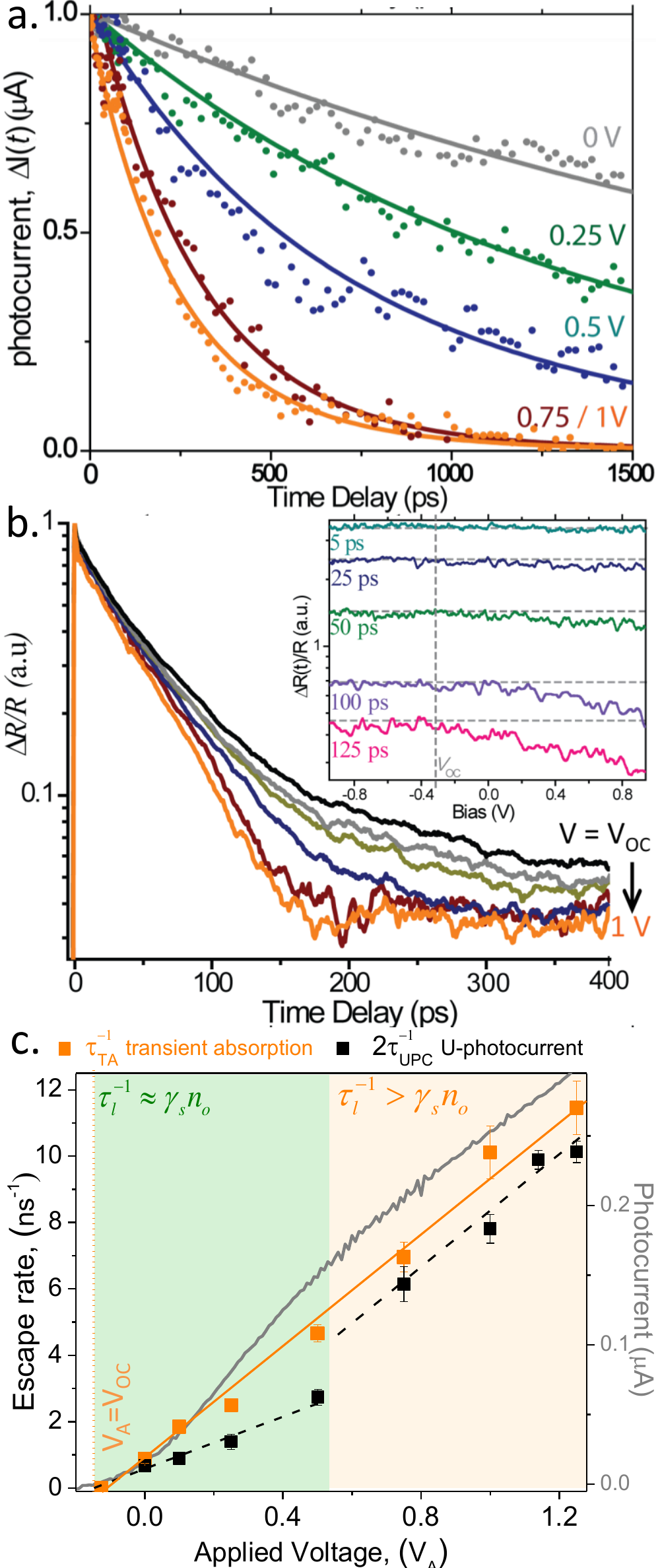}   
        \caption{\textbf{Ultrafast carrier escape, optical vs. PC response.}\textbf{ (a)}  Normalized U-PC response from WSe$_{2}$ accelerate with increasing $V_A$. Solid lines are fits to U-PC response function Eq. 6. \textbf{(b)}  Concurrently measured resonant TA decay rate also accelerate with $V_A$ in the second component. \textit{(inset)} TA  amplitudes decrease  as electrons escape.  \textbf{(c)} The $V_A$-dependent TA  rates (\textit{orange}) and U-PC kinetic rates (\textit{black}) both increase linearly with $V_A$. The device PC-response (\textit{gray line}, right axis) scales likewise. In photoconduction electron drift the full TMD thickness of 56 nm, so rates are plotted as $2\tau^{-1}_{e/h}$. The slope of the linear fits approximate the mobility, $\mu_{e/h}$. }
  \end{figure}   
\textbf{Experimental Analysis II: \textit{ connecting ultrafast   TA and PC electron escape dynamics with device efficiency and response time.}} \\  \indent In section I, the simple  kinetic rate law in Eq. 1 predicted both the photon-flux dependent PC and TA absorption amplitude better than  any phenomenological power-law fits attempted (shown in  supplement S.3).\cite{Massicotte2015,Wang2015} Resolution of   kinetic rates pertinent to the PC response of few-layer TMDs,  require both sub-100 ps time-resolution and selectivity for PC generating dynamic processes. Fig. 2  plots the ultrafast photocurrents (U-PC) concurrently  measured with TA as $V_A$ in increases slowly from $V_{oc}$ to 1.3 V. The results are all repeatable and how minimal impact from the small transport curve hysteresis. This synchronous optical and electronic detection-regime is important to isolate the   ultrafast electron escapes rates from other  recombination  dynamics.   

\indent Prior  works applying U-PC to graphene, carbon nanotubes, and  similar 2D TMDs  suggest U-PC  is predominately sensitive to the competing kinetic rates that limit PC detection  (e.g. extraction vs. recombination).\cite{Graham2013,Graham2013a,Gabor2012,Wang2015,Massicotte2015,Vogt:16}    Figure 2a shows the normalized U-PC kinetics markedly accelerate with increasing applied voltage, consistent with faster carriers extraction. Similar U-PC kinetics on WSe$_2$ photodetectors were first reported by Massicotte \textit{et. al.} and Vogt \textit{et al.} in 2016.\cite{Massicotte2015, Vogt:16}  However, there remains no independently-verified response function that reliably extracts the first-principle kinetic rate constants from the U-PC response. 

Motivated by  the highly predictive PC-response of  Eqn. 3-4 in section I, \ a  related  U-PC response function is obtained by direct piecewise integration of  $n(t)  $  about the delay-time, $t_d$ to give a time-correlated U-PC response function,
\begin{align}
\Delta J_{e/h}(t_d) &=\frac{e}{\tau_{e}}\left(   \int^{t_d}_0 n(t,n_o) dt + \int^{\infty}_{t_d} n(t,n_o+n(t_{d})) dt \right )
\end{align}  
This equation with Eqn. 3 for \textit{n(t)}, simulates the normalized U-PC\ data with only two free-parameters after  the background PC signal, given by   $2\int^\infty_0 n(t,n_o)dt$  is subtracted.  Fig. 2a plots the theoretical U-PC curves after the two-free parameters were optimized, as they capture the kinetic response better than exponential fits commonly to U-PC in the literature.  Nonetheless, the timescales extracted are not radically different than an ad-hoc exponential fit analysis.  All data and simulation are normalized to  peak maximum to compare dynamics (\textit{see supplement section S.4 for examples of uncorrected raw data}). 


     To independently verify the U-PC response function proposed in Eq. 6, we compare the U-PC rates extracted in Fig. 2a to the outwardly much-faster  synchronously obtained  TA kinetics plotted in Fig. 2b. Figure 2b shows the resonantly excited ($K_A$ point) TA microscopy relaxation  increases markedly with applied  voltage ($V_A$), but only in the second decay component.    Figure 2b (\textit{inset}) clearly shows the raw TA signal-amplitude strongly decreases with $V_A$, but only the 75-125 ps range associated with the second decay component.  As  Eq. 5 correctly, the slowest TA relaxation is not observed at $V_A$=0, but instead at the open-circuity voltage, $V_{OC,}$ where  no net-PC is collected. By definition, only TA (not U-PC) can be collected at $V_A=V_{OC}$ when no net carriers are extracted. A least square deconvolution  fit of  the $V_A=V_{OC}$ TA to Eqn. 2  gives $\tau_r =\tau_l\cong92$ ps, and represents the linear recombination rate. Combined with the high (up to 44\%)\ IQE  of the device,  TA kinetics in Fig. 2b   accelerate as the   carrier are extracted from the device at a mean  electronic escape rate  $\tau_{e}^{-1} \cong \tau_l^{-1} - \tau_r^{-1}$. Strikingly in Fig. 2c (\textit{orange squares\textit{)}}, the TA escape rates increase linearly with $V_A$ after the linear recombination rate term is subtracted. Obtained by fitting of  U-PC kinetics   to Eq.  6, Fig. 2c also plots the $\tau^{-1}_{e/h}$ rates obtained from a $L=56$ nm thick WSe$_2$ device. The electron or hole escape rates, $\tau_{ e/h}^{-1}$   in Fig. 2c all increase  with $V_A$ according to  Eqn. 5. The  fits shown yield  slopes further estimate the  electron and hole  mobilities  using Eqn. 5 (see Table 1). 

Interestingly, Fig. 2c  shows that the $E$-field-dependent TA optical escape rates (\textit{orange circles}) agree  well with the  rates obtained from our U-PC response function (\textit{black squares}).  The  agreement of optical and electronic responses   at high $V_A$, suggests   the U-PC response function proposed in Eqn. 6  provides rate-limiting escape kinetics for PC-generation for our TMD devices. 
\begin{table}[ht]
 \begin{tabular}{||l|l|l| c c c| c c | c ||} 
 \hline
 \textit{method}&$V_A$& l  &$\tau_{e/h}$ & $\tau_r$ &$ \tau_l$&$  v_{max}$  &$ \mu_{e/h}$ & $\epsilon_{max}$ \\ [0.5ex] 
  \hline\ TA &1.1&28 & 82 & 92 & 44 & 341 & 0.128& 0.52\\ 
  \hline
$ U-PC_h$ &0.5& 28& 1430 &- & - & 78 & 0.031 & - \\
 \hline
 $U-PC_e$\ &1.1& 56&248 & - & 52 & 245 & 0.121&0.43 \\
  \hline
 PC at 0 V  
 &0&28 & - & -&- & - & - & 0.43\\
 \hline
 PC at 1 V  
 &1&56 & - & -&- & - & - & 0.44\\

   \hline\hline
(\textit{units}) &V&nm & ps  &  & & m/s & cm$^2$/Vs&  \\

 \hline
\end{tabular}
 \caption{\textbf{Table 1.0} - Summary figures of merit. $\mu_{eh}$ are obtained from linear slopes of Fig 2c by Eq. 5.  Both U-PC and TA times shown  independently predict the actual device maximum IQE-PC.}
\end{table}

In the time-domain, PC-generation efficiency($\epsilon$) in Fig. 1b can be approximated by a ratio of competing kinetic rates given by,  $\epsilon \cong \frac{ \tau_{e/h}^{-1}}{\tau_{e/h}^{-1} + \tau_{r}^{-1} +\gamma'n_D n}$. When the defect assisted-Auger scattering rate is small at low photon fluence, $\epsilon_{max}\cong\frac{ \tau_{e/h}^{-1}}{\tau_{e/h}^{-1} + \tau_{r}^{-1} }$.  In Fig. 2c, the WSe$_2$ device intrinsic photocurrent efficiency increases  with applied voltage to yield $\epsilon_{max}=$52\% prediction TA\ kinetics and 43\% from U-PC kinetics. The figures of merit are summarized in Table 1.0, and show the U-PC   $\epsilon_{max}$ matches the actual device performance better than TA. The higher efficiency predicted by of TA is the likely result of overestimating the actual electron drift length, \textit{l} \   as $l \cong L/2$. At high $V_A$ for U-PC, the electrons must traverse the full length of the device for photoconduction so $L$ is known accurately resulting in more accurate efficiency prediction. 

 \begin{figure*}[htbp]
        \includegraphics[height= 1.9 in]{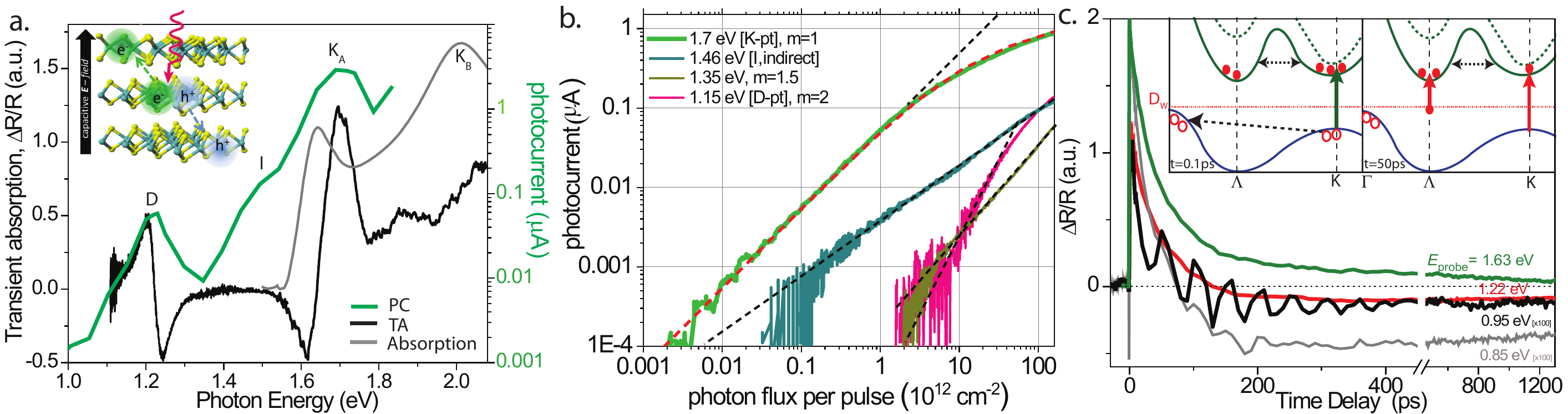}   
\caption{\textbf{Defect mediated carrier relaxation.} \textbf{(a)} Linear  absorption (\textit{gray}), PC (\textit{green}, log-scale)  and TA(\textit{black}) spectrum at $t$=0.5 ps for WSe$_2$. Below the optical bandgap, $K_A$, peaks are   observed at the indirect transition, \textit{I} and defect peak, \textit{D} (\textit{inset}) Interlayer \textit{e-h} extraction. \textbf{(b)} PC-photon flux  dependence of  sub-gap peaks begin linear when excited at $K_A$ but are quadratic  when resonant with the \textit{D}-peak (log-log scaling, dashed black line). At higher-fluxes Auger-recombination accounts for the nonlinearity, and fits well to  Eq. 4  (\textit{red dashed}).  \textbf{(c)}   \textbf{} Sub-gap TA-kinetics for decreasing photon probe energies after an optically resonant pump show a long defect induced ground-state recovery with ns lifetime and an acoustic-oscillation from a weak thermoreflectance response. \textit{\textbf{(inset)}} Band-structure schematic of WSe$_2$ depicting fast  initial carrier thermalization after  resonant pump  (\textit{green arrow}). TA spectral probes (\textit{red arrow}) at the $K_A$(bleach) and $D$ (ESA) peaks preferentially access electron dynamics, whereas  at low $V_A$, U-PC is  predominately sensitive to the slower hole-mobility.  }
\end{figure*}

\subsection{DISCUSSION }    
 
WSe$_2$  in-plane conduction can approach  \textgreater  350 cm$^2$V$^{-1}$s$^{-1}$\cite{kim_enhanced_2016-1,Lee2014}, whereas the out-of-plane conduction is $\sim 10^4$ smaller.  Nonetheless, devices made using the interlayer out-of plane conduction mode are often more efficient than reported in-plane conducting devices.\cite{Lee2014} The combined  large multilayer absorption, short ($\sim$20-30 nm) escape lengths and long recombination times collectively help  TMD achieve high PC-IQE.  However, high defect densities, $n_d$\ common in TMDs can negate this advantage owing to intrinsically slow   electronic carrier  drift velocities measured up to   $\sim$ 250 m/s    (even under high applied voltages, see Table 1.0) where $\tau_e \cong \tau_r$. During interlayer drift, defect-assisted Auger recombination dominates when the incident photon flux is high-enough  ($>$$10^{12}$) enough for carrier scattering (\textit{e-e, e-h, h-h}) to defect sites such as W-vacancies.  \cite{Koperski2015, He2015}

Prior optical ultrafast studies all agree TMDs relaxation dynamics are often dominated by defect-assisted Auger processes. \cite{wang_ultrafast_2015-1, strait_high_2014,Wang2015a,ovesen_interlayer_2019,selig_ultrafast_2019} In 2016, Massicotte \textit{et al.} \cite{Massicotte2015} reported that stacked WSe$_2$ devices can have both $\epsilon>$30\% and fast ($<70$ ps), picosecond electron escape times, $\tau_e$.  Even with our new analysis and the addition of TA, our results and rates obtained still closely match the early seminal works \cite{Massicotte2015,Vogt:16} of U-PC for similar thickness WSe$_2$ \ devices. 
As suggested by spectra  and bandstructure plotted in Fig. 3a and c, a complete description of the relaxation pathways  of resonantly excited WSe$_2 $   must rigorously track transient electron and holes population at all low-lying symmetry points ($\Gamma, \Lambda, K$) and the defect states.  Such complete kinetic models have been the subject of many transient optical absorption and time-resolved photoluminescence studies.\cite{Wang2015a,moody_microsecond_2018,,zhu_enhancement_2018,kaviraj_physics_2019,Pogna2016} Near time zero,  fast Auger processes are indicated by  strongly  square root scaling of the $\Delta R/R$ TA signals on photon flux is  observed.\cite{he_competition_2018}
 After excitation,  the electron-hole pairs dissociate at a rate, $\tau_d^{-1}$ by intervalley themalization or undergo fast-Auger scattering at rate $\gamma_fn_{eh}$ with a kinetic rate of electron-hole pairs written as\cite{Amani2015,Borzda2015,Yu2016,Arora2015,Mouri2014,Rivera2015}, 
\begin{align}
\frac{dn_{eh}}{dt} & =n_{abs} \delta(t) - \frac{n_{eh}}{\tau_{d}}- \frac{n_{eh}}{\tau_{r}}-\frac{1}{2}\gamma_{f}n_{eh}^{2}. \end{align}

\noindent At very short delay times, our TA kinetic decay show signatures of the above kinetic rate law  associated with electronic scattering and dissociation (e.g.   see supplemental section confirming fast photon-flux dependent TA\ kinetics). In particular, at shot delays the initial kinetics accelerate with the incident photon flux.  Fast electron-hole dissociation it is not a rate-limiting process for PC-generation for few-layer WSe$_2$ and so our kinetic treatment of the carrier dynamics after dissociation used in Eq. 1-2 is justified.\cite{moody_exciton_2016,moody_microsecond_2018}


In Fig. 3a the linear absorption (\textit{gray}), TA (\textit{black}) and PC (\textit{green}) spectra  of a WSe$_2$ device are plotted over a 1.0 to 2.1 eV spectral range.    At 1.68 eV both the TA transient bleach and the linear absorption  peak strongly at resonant  \textit{K}$_A$-point transition. Unexpectedly, we also observe a sub-gap excited state absorption (ESA) peak centered at at 1.21 eV.   While such sub-gap peaks are not present in our linear absorption, the  PC-spectrum in Fig. 3a (\textit{green line}, log-scale on right axis) has a matching sub-gap peak visible on semi-log scaling.  The PC spectrum further has a labelled shoulder-peak, \textit{I}  that matches the indirect bandgap of WSe$_2$ at 1.53 eV.\cite{kaviraj_physics_2019}  

To understand the origin of PC and TA response at the labeled D-peak at 1.21 eV, a PC laser flux power-dependence is plotted for each peak-excitation energy in Fig. 3b (log-log scaling).  Upon resonant excitation at $K_A$, the red-dashed line fit to Eq. 4 remarkably captures the PC-response  over five-orders of magnitude and clearly  delineating the onset on of nonlinear-PC response at $\sim$10$^{12}$ photons/cm$^2$ flux. \cite{Sun2014a, Zhao2013}
 For sub-gap region excitation, Fig. 3b shows a PC-response that is superlinear, increasing quadratically with photon flux (slope \textit{m}=2).\cite{zhu_enhancement_2018,massicotte2016}  While direct optical excitation of defect states is generally not possible owing to momentum-matching restriction,  a weakly allowed two-photon excitation is proposed.  Figure 3b supports a two-photon excitation by the characteristic quadratic increase in PC response with photon flux only for the energy matching the ESA TA D-peak in Fig 3a.  Comparison with prior DFT simulations of WSe$_2$ cells of  W metal-metal vacancies suggest a likely origin of this sub-gap defect peak as depicted Fig. 3c.\cite{wu_atomic_2018,moody_microsecond_2018,fang_strong_2014}   

 TA relaxation dynamics shown in Figs. 3-4  vary  greatly depending upon the energetic window of the band-structure probed after resonant excitation.  TA shown in Fig. 2b and 3c are always summation of all electronic kinetic pathways in WSe$_2$ for a selected  probe window.  Fig. 3c (\textit{inset}) schematically shows multi-layer WSe$_2$ bandstructure.  In TA, the holes are largely scattered out of the resonant spectral probe window to the $\Gamma$-point, making resonantly excited TA predominately sensitive to thermalized electron (vs. hole) population.  Conversely, as holes have lower mobility in WSe$_2$, photoextraction is strongly impacted the hole escape rate.  Accordingly, our U-PC escape rates plotted in Fig. 2c  at low applied $V_A$ fits to Eq. 6 and the slope  estimates the out-of-plane hole mobility of 0.031 cm$^{2}$/(Vs).  The hole mobility is  $\sim3-4$x smaller than the electrons, roughly consistent with its larger effective mass as indicated by  DFT of the valence band maximum curvature.\cite{wu_atomic_2018,fang_strong_2014}  At higher $V_A$, the device is predominately in photoconduction mode   and U-PC is instead limited by the photoinduced-electron lifetime.  In Fig. 2c, the slope of  both TA and U-PC escape rate agree giving identical estimates of the electron mobility of 0.128 cm$^{2}$/(Vs).  This time-domain values of the mobility agree reasonably with literature and our transport-based estimate of our device.\cite{moody_exciton_2016,kim_enhanced_2016,fang_strong_2014}

Figure  3c plots the relaxation rates associated with-using sub-gap probe window to access the defect state lifetime.
For sub-gap probe energies below the \textit{D-}peak resonance, the TA\ response is small (\textless100x), and the signal in Fig. 3c at  0.85 and 0.95 eV is predominantly attributed to the long-lived transient thermal response WSe$_2$. This is supported by the incidental presence of a strong TA beat period of 44 ps that further predicts a speed of sound of 2650 m/s in WSe$_2$ which is faster than slower interlayer electronic hopping velocities estimated in Table 1.0.

As previously shown, defect-assisted Auger scattering dominates TA signal above $10^{12}$ photons/cm$^2$. Figure 4a plots the fluence-dependent TA\ kinetics of a resonantly excited 69-layer WSe$_2$ device at such high pump-fluences.   The long ground state recovery component is $\sim$5.6 ns and likely  attributable to defect relaxation and grows in relative amplitude  with photon flux. This coincides with the rate of defect-assisted Auger scattering  growing  with carrier density, \textit{n}. 

\begin{figure}[htbp]
\includegraphics[height=6.6 in]{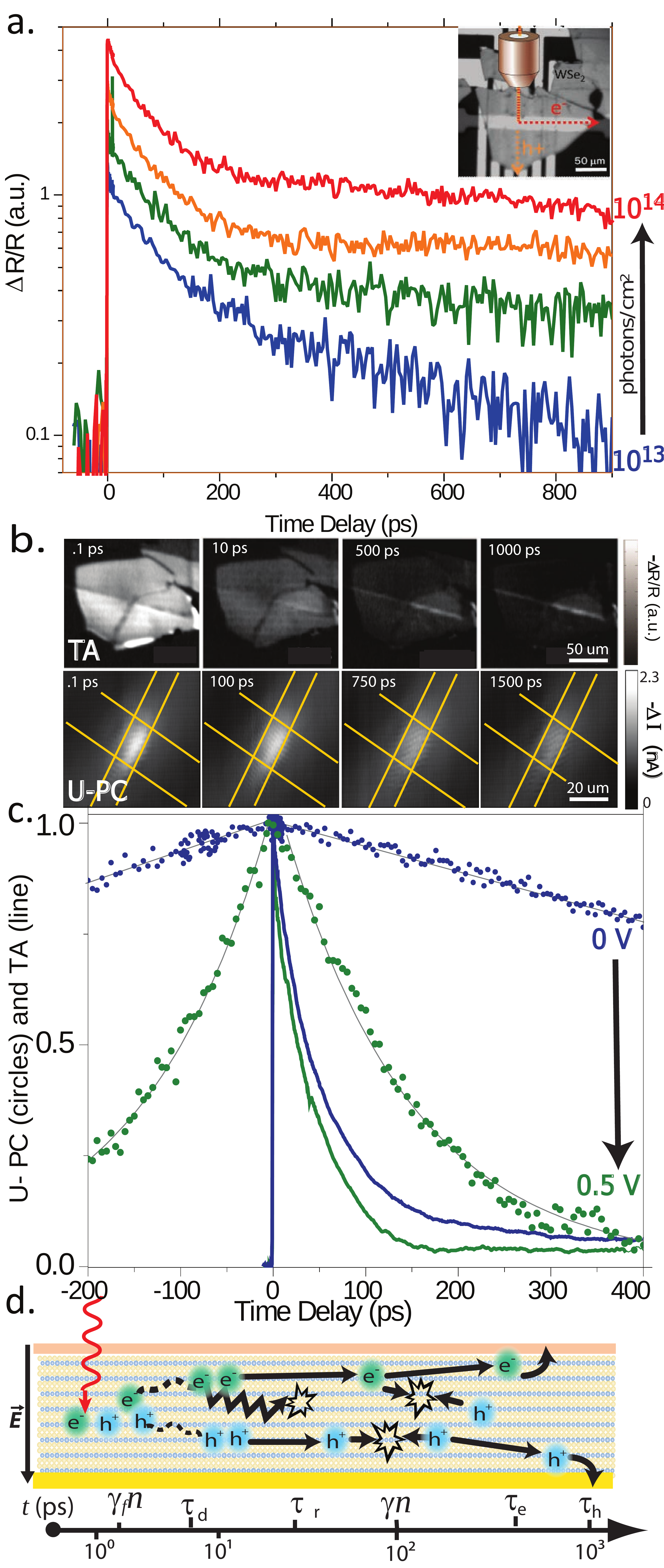}   
\caption{\textbf{TA and PC microscopy suggest a photoextraction time-line. }\textbf{(a)} Long-component kinetic relaxation of the resonantly  probed $K_A$ transition slows as incident photon flux (photons/$cm^{2}$) increase suggesting  an increases defect assisted Auger recombination rate populating long-lived sub-gap  vacancy traps. ($inset$) Ultrafast TA and PC signals are collected simultaneously on the WSe$_2$ device shown.  \textbf{(b)}  TA and U-PC microscopy provide spatial maps of electronic escape and recombination in a WSe$_2$ device. \textbf{(c) }  Summary graphic plots U-PC and TA at $V_A=0$ (\textit{blue}) and 1.5 V (\textit{green}).  Lines are fits to first-principle  kinetic rate laws given by Eq. 2 and 6.   \textbf{(d)}  Cartoon time-line of PC-generation  from \textit{e-h} dissociation to defect-assisted Auger processes. While carriers can escape a 56 nm device in 86 ps, the efficiency is limited by a $\sim$ 90 ps linear recombination even when $\gamma n \cong 0$.}
\end{figure}

Figs. 4b-d summarize our combined TA and U-PC microscopy-measurement response as a function of time, space and applied voltage. A cartoon time-line in Fig. 4d  captures the dominant photocurrent generating  dynamic pathways; from light absorption to photocurrent collection.  The mechanisms suggested in Fig. 1aii anf 4d are supported by simultaneous measurement of both TA and U-PC responses. Taken together, they selectively isolate the PC-relevant kinetic rates of  Auger recombination ($\gamma_{} n(t)$), linear recombination ($\tau_r^{-1}$), and mean escape rate,  $\tau_{e}^{-1}$.   For much thiner, TMD stacks even faster electronic extraction times (down to $\sim$50 ps) are reported  by Massicotte \textit{et al} in similar few-layer WSe$_2$ devices using graphene-based contacts.\cite{Massicotte2015}  Quantitatively similar results are obtained here by using a  first principle kinetic model. While our model is crude to those needed for prior TA studies, we show it is sufficient for predicting PC generation by direct real-time  comparison with TA\ measurements.\cite{Wang2015a,vialla_tuning_2019,li_auger_2018,Yuan2015}
 The agreement between the U-PC rates and  \textit{E}-field-dependent TA rates in Fig. 2c independently validates  that the proposed U-PC response function in Eq. 6   correctly retrieves first principle kinetics.\cite{Palummo2015, Cui2014, bakulin_ultrafast_2016}    \\    \indent

In summary, photodectors of few-layer semiconducting TMDs like WSe$_2$ have low out-of-plane mobilities, and the slow drift velocities make defect-assisted Auger-scattering a   bottleneck to PC-collection even under CW excitation.  Going forward, this combined  TA+U-PC microscopy approach suggests a promising, new way to screen materials for the maximal device PC-collection efficiency by isolating the ultrafast electronic rates that are rate-liming to photocarrier extraction.\cite{bakulin_ultrafast_2016}

 \textbf{\\Conclusions:}

 Recent time-domain studies of stacked WSe$_2$ show TMD devices can be both efficient and fast despite very low out-of-plane mobility.\cite{Vogt:16, Massicotte2015}  Using a first-principle kinetic model involving defect-assisted Auger recombination the nonlinear PC response in Fig. 1c and 3b can be accurately predicted and fit better commonly power-law approximations.  As the applied voltage increases the TA and U-PC extracted escape times increase linearly as $\tau_e \propto |V_A-V_{OC}|^{-1}$, from $\tau_{e/h}=$ 86 ps to 1.6 ns.  Using these rates, a simple kinetic rate model of carrier escape and defect-assisted Auger recombination sufficiently models the device PC-IQE and e nonlinear PC-power dependence over a $10^5$ change in photon flux.  

Matching synchronously obtained TA results in Fig. 2c, provide independent confirmation that the U-PC  first-principle kinetic response function proposed in Eq. 6  analytically extracts the \textit{E}-field-dependent escape rate.       The ratio of competing ultrafast rates extracted from TA and U-PC predict an  $\epsilon_{max}\cong$45\% (PC-IQE), suggesting our TMD\ photodetectors are intrinsically limited by  92 ps recombination time extracted at open-circuit voltage.  \  This $\epsilon_{max}$ closely matches  the 44\% IQE originally measured on our a 56 nm thick  WSe$_2$ devices in Fig. 1b. Both optical and electronic methods gave independently agreeing electronic escape times, and provided out-of-plane electron and hole mobilities of  0.129 and 0.031 cm$^2$/V$s$ respectively for WSe$_2$. The large absorption coefficient, short escape lengths and the suppressed recombination of carriers collectively explain why WSe$_2$  -based photosensors can\ still be both fast ($<100$ ps) and efficient despite low out-of-plane carrier mobility and strong Auger recombination process. By combining $E$-field dependent ultrafast photocurrent with transient absorption microscopy we idenified the dominant kinetic bottlenecks inhibiting photocurrent efficiency for stacked few-layer TMD semiconductors phtodetectors. This approach may further help indentify the maximum quantum efficiency in other promising materials where electron extraction is kinetically favorable over recombination.

\section{Method Details:} 
   \textbf{\\ Device preparation:} 150 nm thick gold  contact pad are patterned on a silicon wafer. Both WSe$_2$ and  MoS$_2$ was mechanically exfoliated and inspected optically to identify  thick that was confirmed by atomic force microscopy (AFM). Optimal thicknesses ranged from 50 to 100 layers such that $>80$\% of incident light is  absorbed, but the mean path for electron escape is short enough to efficiently collect.\cite{Yu2013,huang_lateral_2014} TMD samples were then spun-cast with MMA polymer, transferred mechanically onto the gold electrodes and released  thermally after careful cleaning steps optimize contacts.  Finally, an optically translucent top gate of titanium/gold (2nm/16 nm) was deposited along the top of the exfoliated TMD material. Eight independently working devices with strong, uniform PC response as shown in Fig. 1a were studied, and all gave similar result when the stack-layer thickness was accounted for. 

\textbf{Ultrafast photocurrents and transient absorption microscopy:} After  resonant optical pumping of the $K_A$-point transition shown in Fig. 3c, the transient spectra and  kinetics are measured using confocal scanning U-PC and TA microscopy.\cite{hartland2010}  Collinear pump-and probe pulses were obtained from two independently tunable outputs of an ultrafast Ti:Sapphire  oscillator (Coherent Chameleon Ultra II, 80 MHz) pumping an optical parametric oscillator (APE-Compact).  Spectrally resolved TA spectral measurements resonantly pump the \textit{K}-point transition resonance at 780 nm, and a white-light supercontinuum probe was used to capture both transient spectra.  Cross-correlation of the pump and probe after the objective yielded a  pulse duration of about 160 fs.     
    
    After a mechanical delay stage, both the pump and the probe beams were aligned in a collinear geometry, raster-scanned by piezo-scanning mirror and coupled into a confocal scanning microscope via a 50X IR-region enhanced, achromatic objective (NA= 0.65).  Transient absorption signals were detected by measuring the probe beam on with a TE cooled InGaAs detector connected to a Zurich HF2-LI lock-in amplifier with current preamplifier. The pump beam was modulated at either a 25 kHz using an acousto-optical (AO-modulator (Gooch \& Housego) to enable high-frequency lock-in detection of the differential reflectivity, $\Delta R/R$. Frequency was swept to ensure beam modulation frequency is independent of the photoresponse measured.  Appropriate optical filters or Acton monochromator were used in front of the detector to block the pump beam and select probe spectral window. The pump and probe spot sizes on the sample were  determined to $\sim$1.5 $\mu$m, by fitting to a confocal scanning reflection profile of lithographic gold pad edges.  All the measurements were done at 295 K. The probe power was  fixed at ($\sim 0.5 \times 10^{12} $ photons/cm$^2$ ) for the pump power dependence measurements.

     The  steady-state scanning photocurrent geometry discussed above is modified to accommodate simultaneous transient absorption and U-PC measurements.  Specifically, a linear delay stage is added to control the relative timing of the two incident laser pulses (one at 780 nm and one 810 nm to present pulse optical  interferometry) which are aligned in a collinear geometry before a piezo scanning mirror (PI, $\# S-334.2SL$) and coupled into the microscope (Olympus BX-61WI).  Both U-PC and TA microscopy signals are collected as a function of delay-time using lock-in amplifiers and amplitude-modulation the two beam lines shown in Fig. 1a.  The ultrafast  photoresponse was demodulated by either single or difference frequency chopping mode.  Photon-flux dependent measurements used a motorized polarizer-waveplate combination to study the TA and PC response over a $\sim 10^5$ continuous range.

\textbf{Acknowledgements}: 
This work acknowledges the Spectroscopy Society of Pittsburgh, NSF MRI grant 1920368  and Oregon Best for project support.  S.-F. Shi acknowledges support by AFOSR\ through grant No. FA9550-18-1-0321

\bibliography{shortbib2}

\end{document}